# Critique of optical transition theories based on projection and population criteria

Nam Lyong Kang

*Department of Nanomechatronics Engineering,*

*Pusan National University, Miryang 627-706, Republic of Korea*

## Abstract

Some many-body theories of optical transitions in solids were examined from projection and population criteria. The results showed that state-independent projection methods cannot be applied to electron systems with non-uniform energy spectra. Moreover, neglecting some important terms in a second-order approximation leads to invalidation of the population criterion. In addition, a valid theory satisfying these two criteria can be obtained using a proper state-dependent projection operator and Kang-Choi reduction identity, and the result can be interpreted using a diagram, which can model the quantum dynamics of electron in solids.

## I. INTRODUCTION

The rapidly growing development of solid state technology has prompted increasing interest in the properties of electron in solids interacting with a background. Although many theoretical approaches for this problem have been reported [1-5], this study focused on the many-body projection technique because well-defined projectors can yield elegant formalism [6-10]. Nevertheless, there is some controversy regarding the validity and advantages of these theories. A proper theory should satisfy some of the many-body criteria. The factors involved in the theories of optical transitions in electron-phonon systems include the energy denominator factor enforcing energy conservation, the population factor combining the electron and phonon distribution functions, and the interaction factor describing electron-phonon coupling. In addition, the conservation of momentum needs to be considered in a many-body calculation.

This paper introduces two criteria for assessing the validity of many-body theories based on the projection technique, the population criterion and the projection criterion. The "population criterion" states that the electron and phonon distribution functions should be combined in multiplicative forms [see Eq. (27) in Ref. 8 and Eq. (4.10) in Ref. 9] because electrons (fermions) and phonons (bosons) belong to different categories in the quantum-statistical classification. This criterion is reflected in the population factor in a proper combination with the energy denominator factor. This population factor should be combined properly with the interaction factor, which interprets the occurrence of local fluctuations due to electron-phonon interactions in such a way that a transition can occur via implicit states. On the other hand, the other projection methods [3-7] do not satisfy this criterion because two distribution functions are simply added (or subtracted), such as $N_q + 1/2 \mp 1/2 \pm f_\alpha$, where $N_q$ and $f_\alpha$ are the Planck distribution function for a phonon with the wave vector $q$ and the Fermi distribution function for an electron with energy $E_\alpha$, respectively.

The second criterion, the "projection criterion", states that the basis (or projection axis) for the projection operator should be chosen in such a way that for the diagonal part ($L_\mathrm{d}$) of the Liouville operator $L$ corresponding to the equilibrium Hamiltonian $H$, $L_\mathrm{d} \times$ (basis) is proportional to (basis). Here $L = L_\mathrm{d} + L_\mathrm{v}$ and $L_\mathrm{d} = L_\mathrm{e} + L_\mathrm{p}$, where $L_\mathrm{e}$, $L_\mathrm{p}$, and $L_\mathrm{v}$, denote the Liouville operators corresponding to the unperturbed electron Hamiltonian $H_\mathrm{e}$, phonon

Hamiltonian $H_\text{p}$, and electron-phonon interaction Hamiltonian $V$, respectively. In most cases, the many-body current density operator, $J^+ = J_x + iJ_y$, is chosen as the basis of the projection operator [see Eq. (2.28) in Ref. 6, Eq. (3.11) in Ref. 7, and Eq. (11) in Ref. 8]. This choice can deal with the optical transitions in systems, whose energy spectra are uniform, a typical example being the cyclotron transition in materials with a symmetrical band structure. Note that $L_\text{d}J^+ = \hbar\omega_\text{c} J^+$ for a cyclotron transition, where $\omega_\text{c}$ is the cyclotron frequency. This choice, however, cannot be used for systems with non-uniform energy spectra because $L_\text{d}J_z$ is not proportional to $J_z$.

This paper discusses the validity of a state-dependent projection operator (SDPO) and the Kang-Choi reduction identity (KCRI) method [8-10] in dealing with the optical transition in the general system including those with non-uniform energy spectra. An additional aim of this study was also to show how a physically acceptable form of optical conductivity can be obtained by combining the SDPO and KCRI, and the line shape function in the optical conductivity can be interpreted by a diagram, which can provide a physical intuition for quantum dynamics of an electron in a solid.

## II. REVIEW OF THEORIES FOR CYCLOTRON TRANSITION

For an incident electromagnetic wave of angular frequency $\omega$ circularly polarized in the $xy$ plane, the optical conductivity tensor of an electron system in solids is given as [5-8]

$$\sigma_{+-}(\omega) = \frac{i}{\omega_\text{c}} \sum_\alpha (j_\alpha^+)^* <(\hbar\bar{\omega} - L)^{-1} J^+>_\alpha. \tag{1}$$

where $\bar{\omega} \equiv \omega - ia (a \to 0^+)$, $j^\pm \equiv j_x \pm ij_y$ for a single electron current density operator $j$, $X_\alpha \equiv <\alpha+1|X|\alpha>$, and

$$<X>_\alpha \equiv \text{Tr}\{\rho(H)[X, a_{\alpha+1}^+ a_\alpha]\}, \tag{2}$$

where Tr denotes the many-body trace, $\rho(H)$ is the equilibrium density operator, $[A, B]$ is the commutator, and $a_\alpha^+ (a_\alpha)$ denotes the creation (annihilation) operator for an electron in the state $|\alpha>$ with energy $E_\alpha$.

This paper considers a system of electrons interacting with phonons under a time-independent magnetic field $B$ applied along the $z$ axis. The Hamiltonian is given by

$$H_{\rm e} + H_{\rm p} = \sum_{\alpha} E_{\alpha} a_{\alpha}^{+} a_{\alpha} + \sum_{q} \hbar\omega_{q}\left(b_{q}^{+} b_{q} + 1/2\right), \tag{3}$$

$$V = \sum_{\alpha,\beta} \sum_{q} C_{\alpha\beta}(q) a_{\alpha}^{+} a_{\beta}\left(b_{q} + b_{-q}^{+}\right). \tag{4}$$

Here $b_q^+(b_q)$ denotes the creation (annihilation) operator for a phonon with energy $\hbar\omega_q$ and $C_{\alpha\beta}(q) = V_q <\alpha|\exp(i\mathbf{q}\cdot\mathbf{r})|\beta>$ is the electron-phonon interaction matrix element, where $V_q$ is the coupling factor that depends on the mode of phonons.

## A. Theory of projection − isolation method

To calculate Eq. (1), Badjou and Argyres [6] introduced projection operators for an arbitrary operator $Y$ defined as

$$PY \equiv J^{+} <Y>_{\alpha}/<J^{+}>_{\alpha}, \quad P' \equiv 1 - P, \tag{5}$$

which is called state-independent because the projection basis, $J^+$, is state-independent. After some systematic calculations, they obtained [see Eq. (2.32) in Ref. 6]

$$<(\hbar\bar{\omega} - L)^{-1}J^{+}>_{\alpha} = \frac{<J^{+}>_{\alpha}}{\hbar\bar{\omega} - <(L + LG'(\bar{\omega})P'L)J^{+}>_{\alpha}/<J^{+}>_{\alpha}}, \tag{6}$$

where $G'(\bar{\omega}) \equiv (\hbar\bar{\omega} - P'L)^{-1}$. Using $L_{\rm d}J^{+} = \hbar\omega_{\rm c}J^{+}$,

$$P'L_{\rm d}J^{+} = 0, \tag{7}$$

and $<L_{\rm v}J^{+}>_{\alpha} = 0$, they obtained

$$<LG'(\bar{\omega})P'LJ^{+}>_{\alpha} = <L_{\rm v}G'(\bar{\omega})P'L_{\rm v}J^{+}>_{\alpha} \tag{8}$$

in the second order approximation in $L_{\rm v}$. They neglected the term, $<L_{\rm d}G'(\bar{\omega})P'L_{\rm v}J^{+}>_{\alpha}$, in Eq. (8) [see Eq. (2.32) in Ref. 6] assuming that $<L_{\rm d}P'Y>_{\alpha} = 0$ but this should be considered because $<L_{\rm d}P'Y>_{\alpha} = <L_{\rm d}Y>_{\alpha} - <L_{\rm d}PY>_{\alpha} = <L_{\rm d}Y>_{\alpha} - \hbar\omega_{\rm c} <Y>_{\alpha} \neq 0$. Their result for the electron system interacting with phonons contains terms, such as $N_q + 1/2 \pm 1/2 \mp f_{\alpha}$, which does not satisfy the population criterion. If the missing term is included properly, the results may satisfy the criterion.

## B. Theory of isolation – projection method

Applying the isolation operator first, i.e., using the isolation-projection method, Cho and Choi obtained [7]

$$< (\hbar\bar{\omega} - L)^{-1}J^+ >_\alpha = \frac{i<J^+>_\alpha}{i\hbar(\bar{\omega} - \omega_c) + \Gamma_\alpha(\bar{\omega})}. \tag{9}$$

Here,

$$\Gamma_\alpha(\bar{\omega}) \equiv \frac{<[Q(\bar{\omega}) + Q(\bar{\omega})G''(\bar{\omega})P'Q(\bar{\omega})]J^+>_\alpha}{i<J^+>_\alpha}, \tag{10}$$

where $Q(\bar{\omega}) \equiv L_v\Delta + L_v G'(\bar{\omega})\Delta' L_v\Delta$, $\Delta$ and $\Delta'$ are the isolation operators, $G'(\bar{\omega}) \equiv (\hbar\bar{\omega} - L_d - \Delta' L_v)^{-1}$, and $G''(\bar{\omega}) \equiv [\hbar\bar{\omega} - L_d - P'Q(\bar{\omega})]^{-1}$. They neglected the second term in the numerator in Eq. (10) [see Eq. (3.21) in Ref. 7] in the second order approximation in $L_v$, but this should be considered because it contains the second order terms of $L_v$, choosing $Q(\bar{\omega}) \approx L_v\Delta$ and $G''(\bar{\omega}) \approx (\hbar\bar{\omega} - L_d)^{-1}$. Their result for the electron system interacting with phonons also contains the terms, $N_q + 1/2 \pm 1/2 \mp f_\alpha$, so it does not satisfy the population criterion. The result may satisfy the criterion if the missing term is included properly.

## C. Theory of projection – reduction method

Eq. (1) can be calculated using only the projection operator given in Eq. (5). The result is given by [Eq. (15) in Ref. 8]

$$\sigma_{+-}(\omega) = \frac{i}{\hbar\omega}\sum_\alpha (j_\alpha^+)^* \frac{<J^+>_\alpha}{\bar{\omega} - \omega_c - \Gamma_\alpha(\bar{\omega})} \tag{11}$$

in the second-order approximation in $L_v$, where

$$\Gamma_\alpha(\bar{\omega})\hbar <J^+>_\alpha \equiv <LP'(\hbar\bar{\omega} - LP')^{-1}L_v J^+>_\alpha. \tag{12}$$

To calculate Eq. (12) without missing terms, the old version of the KCRI was introduced as follows [see Eq. (18) in Ref. 8]:

$$\begin{aligned}&\text{Tr}\{\rho(H)[LP'X, a_\alpha^+ a_{\alpha+1}]\}\\ &= \text{Tr}\{\rho(H)[L_v a_\alpha^+ a_{\alpha+1}, X]\} - \text{Tr}\{\rho(H)[L_v PX, a_\alpha^+ a_{\alpha+1}]\},\end{aligned} \tag{13}$$

where the fact that $\rho(H)$ commutes with $H$ and $\text{Tr}(ABC) = \text{Tr}(BCA)$ was considered. The results for an electron system interacting with phonons contains the terms $(N_q + 1/2 \pm 1/2)f_\alpha(1 - f_\beta)$, so it satisfies the population criterion [see Eq. (27) in Ref. 8].

### III. THEORY FOR GENERAL OPTICAL TRANSITION

When an electromagnetic wave of frequency $\omega$ is applied to a system in the $j$-direction, the linear optical conductivity in the dipole approximation can be expressed as [9]

$$\sigma_{ij}(\omega) = -e \sum_{\gamma,\delta} (r_j)_{\gamma\delta} \text{Tr}\{\rho(H)[(\hbar\bar\omega - L)^{-1}J_i, a_\gamma^+ a_\delta]\}, \tag{14}$$

where $J_i$ is the $i(=x,y,z)$ component of the many electron current density operator, $J$. If state-independent projection operators such as Eq. (5) are used, Eq. (7) is not satisfied because $L_d J_i \neq$ c-number$\times J_i$, so Eq. (14) cannot be calculated further.

On the other hand, using the second quantized form of $J_i$, $J_i = \sum_{\alpha,\beta} (j_i)_{\alpha\beta} a_\alpha^+ a_\beta$, Eq. (14) becomes

$$\sigma_{ij}(\omega) = -e \sum_{\alpha,\beta} \sum_{\gamma,\delta} (j_i)_{\alpha\beta}(r_j)_{\gamma\delta} \text{Tr}\{\rho(H)[(\hbar\bar\omega - L)^{-1} a_\alpha^+ a_\beta, a_\gamma^+ a_\delta]\}, \tag{15}$$

which can be calculated using the state-dependent projection operators defined as follows:

$$PX \equiv a_\alpha^+ a_\beta <X>_{\gamma\delta} / <a_\alpha^+ a_\beta>_{\gamma\delta}, \quad Q = 1 - P \tag{16}$$

for an arbitrary operator $X$, where

$$<X>_{\gamma\delta} \equiv \text{Tr}\{\rho(H)[X, a_\gamma^+ a_\delta]\}. \tag{17}$$

Note that Eq. (7) is satisfied because $L_d a_\alpha^+ a_\beta = (E_\alpha - E_\beta) a_\alpha^+ a_\beta$ and $Q a_\alpha^+ a_\beta = 0$.

Applying the identity $1 = P + Q$ on the right-hand side of the Liouville operator $L$ in Eq. (15) as $L = L(P + Q)$ results in [see Eq. (4.7) in Ref. 9]

$$\sigma_{ij}(\omega) = -e \sum_{\alpha,\beta} \sum_{\gamma,\delta} \frac{(j_i)_{\alpha\beta}(r_j)_{\beta\alpha}(f_\alpha - f_\beta)}{\hbar\bar\omega - (E_\alpha - E_\beta) - \Gamma_{\alpha\beta}(\bar\omega)}, \tag{18}$$

where

$$\Gamma_{\alpha\beta}(\bar\omega)(f_\alpha - f_\beta) = \text{Tr}\{\rho(H)[L(\hbar\bar\omega - LQ)^{-1} L_v a_\alpha^+ a_\beta, a_\beta^+ a_\alpha]\} \tag{19}$$

which can be calculated using the KCRI given as

$$\text{Tr}\{\rho(H)[LX, A]\} = -\text{Tr}\{\rho(H)[X, LA]\}. \tag{20}$$

The result contains the terms $(N_q + 1/2 \pm 1/2)f_\alpha(1 - f_\beta)$, so it satisfies the population criterion [see Eq. (21)].

## IV. COMPARISON OF THE THEORIES

Table I shows whether the projection criterion is satisfied for several theories. The theories using the state-independent basis, Refs. 6-8, can be only used for a cyclotron transition because $L_d \times$(basis)$\neq$c-numver$\times$(basis) for the non-uniform energy spectra, but the theory using the state-dependent basis, Ref. 9, can always be applied because the projection criterion is always satisfied by $L_d a_\alpha^+ a_\beta = (E_\alpha - E_\beta) a_\alpha^+ a_\beta$.

TABLE I: Projection criterion.

| Theory | Basis | Application | |
|---|---|---|---|
| | | Cyclotron transition | General optical transition |
| Refs. 6-8 | $J$ (state independent) | $L_d J^+ = \hbar \omega_c J^+$ (applicable) | $L_d J_i \neq$c-number$\times J_i$ (not applicable) |
| Ref. 9 | $a_\alpha^+ a_\beta$ (state dependent) | $L_d a_{\alpha+1}^+ a_\alpha = \hbar \omega_c a_{\alpha+1}^+ a_\alpha$ (applicable) | $L_d a_\alpha^+ a_\beta = (E_\alpha - E_\beta) a_\alpha^+ a_\beta$ (applicable) |

Table II shows whether the population criterion is satisfied in the second-order approximation in $L_v$. Ref. 6 contains a missing term, $L_d G'(\overline{\omega}) P' L_v J^+$, in Eq. (2.32), and Ref. 7 also has a missing term, $Q(\overline{\omega}) G''(\overline{\omega}) P' Q(\overline{\omega}) J^+$, in Eq. (3.19). Therefore, they do not satisfy the population criterion. They might satisfy the population criterion if the missing terms are properly included. On the other hand, Refs. 8 and 9 do not have the missing terms and satisfy the criterion.

TABLE II: Population criterion.

| Theory | Population combination (population criterion) | Missing term |
|---|---|---|
| Ref. 6 | $N_q + 1/2 \mp 1/2 \pm f_\alpha$ (unsatisfied) | $L_d G'(\overline{\omega}) P' L_v J^+$ in Eq. (2.32) |
| Ref. 7 | $N_q + 1/2 \mp 1/2 \pm f_\alpha$ (unsatisfied) | $Q(\overline{\omega}) G''(\overline{\omega}) P' Q(\overline{\omega}) J^+$ in Eq. (3.19) |
| Refs. 8, 9 | $(N_q + 1/2 \pm 1/2) f_\alpha (1 - f_\beta)$ (satisfied) | There is no missing term |

## V. VALIDITY OF THE SDPO AND KCRI METHOD

A formula satisfying the two criteria can be obtained using the SDPO and KCRI. The line shape function for a system of electrons interacting with phonons can be obtained from Eqs. (19) and (20) as follows [see Eq. (4.8) in Ref. 9]:

$$\Gamma_{\alpha\beta}(\overline{\omega})(f_\alpha - f_\beta)$$
$$= \sum_{q,\lambda} \left[ \tilde{C}_{\alpha\lambda}(q) \{ G_{\lambda\beta}(+\omega_q) P_+(\lambda, \beta) + G_{\lambda\beta}(-\omega_q) P_-(\lambda, \beta) \} \right.$$
$$\left. + \{ G_{\alpha\lambda}(+\omega_q) P_+(\alpha, \lambda) + G_{\alpha\lambda}(-\omega_q) P_-(\alpha, \lambda) \} \tilde{C}_{\lambda\beta}(q) \right]$$
$$= A + B + C + D. \tag{21}$$

Here, the electron-phonon interaction coupling factor (C-factor), $\tilde{C}_{\alpha\beta}(q)$, is defined as

$$\tilde{C}_{\alpha\beta}(q) \equiv |V_q < \alpha|\exp(i\mathbf{q} \cdot \mathbf{r})|\beta >|^2 \tag{22}$$

by which the momentum conservation is satisfied. In Eq. (21), the energy denominator factors (G-factor), $G_{\alpha\beta}(\pm\omega_q)$, can be defined as

$$G_{\alpha\beta}(\pm\omega_q) \equiv \delta(\hbar\omega + E_\alpha - E_\beta \mp \hbar\omega_q) \tag{23}$$

by which the energy conservation is maintained, i.e., $E_\alpha + \hbar\omega = E_\beta \pm \hbar\omega_q$, and the population factors (P-factor), $P_{\alpha\beta}(\alpha, \beta)$, are defined as follows:

$$P_+(\alpha,\beta) \equiv (N_q + 1)f_\alpha(1 - f_\beta) - N_q f_\beta(1 - f_\alpha) \qquad (24)$$

$$P_-(\alpha,\beta) \equiv N_q f_\alpha(1 - f_\beta) - (N_q + 1)f_\beta(1 - f_\alpha) \qquad (25)$$

by which the population criterion is satisfied and a diagrammatic interpretation is possible.

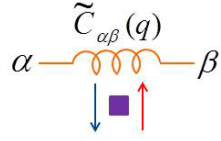

**FIG. 1.** Diagrammatic representation of the interaction factor, $\tilde{C}_{\alpha\beta}(q)$. The inward (red) and outward (blue) arrows, respectively, denote the emission and absorption of a phonon with wave vector $q$.

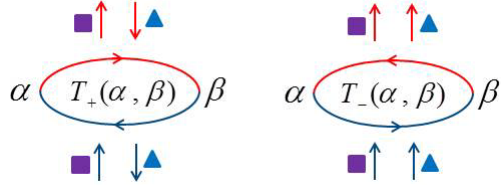

**FIG. 2.** Diagrammatic representation of the transition factors, $T_\pm(\alpha,\beta)$. $T_+(\alpha,\beta)$ and $T_-(\alpha,\beta)$ correspond to clockwise and counterclockwise loops, respectively. The upper (red) and lower (blue) half circles correspond to the phonon emission and absorption processes, and the forward and backward processes correspond to the photon absorption and emission processes, respectively.

For a diagrammatic interpretation, the green circle, blue triangle and purple square denote the electron, photon and phonon, respectively. $\tilde{C}_{\alpha\beta}(q)$ is represented by a spring, which absorbs (inward red arrow) or emits (outward blue arrow) a phonon with a wave vector $q$ (Fig. 1). The transition factors, $T_\pm(\alpha,\beta)$, are defined as

$$T_\pm(\alpha,\beta) \equiv G_{\alpha\beta}(\pm\omega_q)P_\pm(\alpha,\beta), \qquad (26)$$

where $T_+(\alpha,\beta)$ and $T_-(\alpha,\beta)$ correspond to the clockwise and counterclockwise loops, respectively (Fig. 2). The upper (red) and lower (blue) half circles correspond to phonon emission and absorption processes, and the forward and backward processes correspond to the photon absorption and emission processes, respectively.

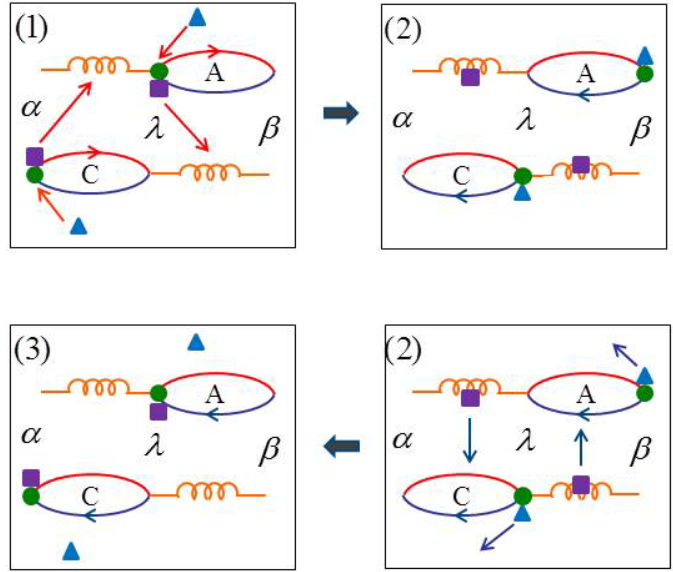

**FIG. 3.** Diagrammatic interpretation of processes A and C in Eq. (21). The green circle, blue triangle, and purple square denote the electron, photon, and phonon, respectively.

Figure 3 shows a diagrammatic interpretation of the first (A) and third (C) terms in Eq. (21). The physical meaning of the first term (A) in Eq. (21) or Fig. 3 is as follows. In the process from stage (1) to stage (2), $\tilde{C}_{\alpha\lambda}(q)$ means that an implicit state, $\lambda$, is coupled with an initial state, $\alpha$, by a phonon with a wave vector $q$ and $T_+(\lambda,\beta)$ means that an electron transits from the implicit state to a final state, $\beta$, absorbing a photon and emitting a phonon with wave vector, $q$, to the spring. The $\lambda$ state is called the implicit state because it is contained in the line shape function [Eq. (19)], not in the conductivity tensor [Eq. (18)]. An electron transits from the final state to the implicit state emitting a photon and absorbing a phonon emitted from the spring in the reverse process from stages (2) to (3). The electron transition process forms a loop because phonon absorption and emission processes maintain a balance. A net transition is possible because the combinations of distribution functions in Eqs. (24) and (25) are different, and the energies of the implicit states are determined by the energies of the final (or initial) state, photon and phonon. Processes C, B and D in Eq. (21) can be interpreted by diagrams in a similar manner.

## VI. CONCLUSION

This paper showed that a theory, which is generally applicable and satisfies the population criterion, could be obtained using a state-dependent projection operator (SDPO) and the Kang-Choi reduction identity (KCRI). The result satisfying the population criterion could be interpreted by a diagram, which can model the quantum dynamics of an electron in solids. Recently, a new magnetic susceptibility formula [11, 12] satisfying the two criteria and including the spin flip and conserving processes properly was derived using the SDPO and KCRI. Understanding how the distribution functions are included is important because the temperature dependence of the spin relaxation time is caused by the Planck and Fermi distribution functions. Therefore, proper theories for the response phenomena should be derived using the SDPO with the KCRI. The SDPO and KCRI method are expected to be applied to other optical transition phenomena in solids.